\documentclass[12pt]{article}
\usepackage{graphicx}
\usepackage{amssymb}

 \def\be{\begin{equation}}
 \def\ee{\end{equation}}
 \def\bea{\begin{eqnarray}}
 \def\eea{\end{eqnarray}}

\def\2{\frac{1}{2}}
\def\4{\frac{1}{4}}

\def\np#1{{\sl Nucl.~Phys.~\bf B#1}}
\def\pl#1{{\sl Phys.~Lett.~\bf B#1}}
\def\pr#1{{\sl Phys.~Rev.~\bf D#1}}
\def\prl#1{{\sl Phys.~Rev. Lett.~\bf #1}}

\def\cqg#1{{\sl Class.~Quant.~Grav.~\bf #1}}

\catcode`\@=11
\def\@citex[#1]#2{%
\if@filesw \immediate \write \@auxout {\string \citation {#2}}\fi
\@tempcntb\m@ne \let\@h@ld\relax \def\@citea{}%
\@cite{%
  \@for \@citeb:=#2\do {%
    \@ifundefined {b@\@citeb}%
      {\@h@ld\@citea\@tempcntb\m@ne{\bf ?}%
      \@warning {Citation `\@citeb ' on page \thepage \space undefined}}%
      {\@tempcnta\@tempcntb \advance\@tempcnta\@ne%
      \@tempcntb\number\csname b@\@citeb \endcsname \relax%
      \ifnum\@tempcnta=\@tempcntb 
        \ifx\@h@ld\relax%
          \edef \@h@ld{\@citea\csname b@\@citeb\endcsname}%
        \else%
          \edef\@h@ld{\ifmmode{-}\else--\fi\csname b@\@citeb\endcsname}%
        \fi%
      \else
        \@h@ld\@citea\csname b@\@citeb \endcsname%
        \let\@h@ld\relax%
      \fi}%
    \def\@citea{,\penalty\@highpenalty\,}%
  }\@h@ld
}{#1}}

\def\@citeb#1#2{{[#1]\if@tempswa , #2\fi}}
\def\@citeu#1#2{{$^{#1}$\if@tempswa , #2\fi }}
\def\@citep#1#2{{#1\if@tempswa , #2\fi}}

\title{Perturbative calculation of quasi-normal modes of arbitrary spin in Schwarzschild spacetime}
\author{{\bf Suphot Musiri}\footnote{suphot@swu.ac.th}\\
\em Department of Physics,
Srinakharinwirot University, \\
\em Bangkok 10110, Thailand. \\ \\
\and {\bf George Siopsis}\footnote{siopsis@tennessee.edu}\\
\em Department of Physics and Astronomy,
The University of Tennessee, \\
\em Knoxville, TN 37996 - 1200, USA.
}
\date{March 2007}
\begin{document}

\maketitle
\vspace{-5in}\hfill UTHET-06-0701\vspace{5in}

\abstract{We calculate analytically the asymptotic form of quasi-normal modes of perturbations of arbitrary spin
of a Schwarzschild black hole including first-order corrections.
We use the Teukolsky equation which applies to both bosonic and fermionic modes.
Remarkably, we arrive at explicit expressions which coincide with those derived using the Regge-Wheeler equation for integer spin.
Our zeroth-order expressions agree with the results of WKB analysis.
In the case of Dirac fermions, our results are in good agreement with numerical data.
}
\newpage


Quasi-normal modes govern the response of a black hole to external perturbations.
In general, they possess a spectrum of complex frequencies due to the leakage of information into the horizon. Their observation will reveal information about the characteristics of the black hole.
In asymptotically AdS spaces, they are related to properties of the dual conformal field theories on the boundary through the AdS/CFT correspondence.
In asymptotically flat spaces, apart from observational possibilities, interest in QNMs has arisen~\cite{bibx1,bibx2,bibx3,bibx4,bibx5,bibx6,bibx7,bibx8,bibx9,bibx10,bibx11} because the asymptotic form of the spectrum was shown to be related to the Barbero-Immirzi parameter~\cite{bibbi1,bibbi2} of Loop Quantum Gravity~\cite{biblqg1,biblqg2,biblqg3,biblqg4,biblqg5}.
The asymptotic form of the spectrum normalized by the Hawking temperature is given by
\be\label{eq1} \frac{\omega_n}{T_H} \approx -(2n+1)\pi i + \ln 3 \ee
for scalar and gravitational perturbations. This has been derived numerically~\cite{bibn1,bibn2,bibn3,bibn4,bibn5}
and subsequently confirmed analytically~\cite{bibx3,bibx5}.
The analytical value of the real part was first conjectured by Hod~\cite{bibhod} based on the form of the horizon area spectrum proposed by Bekenstein and Mukhanov~\cite{bibbm}.
Its value is intriguing in Loop Quantum Gravity, suggesting that the gauge group should be $SO(3)$ instead of the expected $SU(2)$, as the latter would lead to $\Re\omega/T_H \approx \ln 2$ asymptotically.

The asymptotic expression~(\ref{eq1}) was analytically derived and generalized to arbitrary {\em integer} spin $j$~\cite{bibx5}
\be\label{eq1j} \frac{\omega_n}{T_H} \approx -(2n+1)\pi i + \ln (1+2\cos\pi j) \ee
and a perturbative expansion
was established~\cite{bibx11}.
Extending these analytical methods to half-integer spin (such as the Dirac field) is not straightforward.
Dirac quasi-normal frequencies have been calculated numerically~\cite{bibD1,bibD2,bibD3,bibD4,bibD5,bibD6,bibD7,bibD8,bibD9,bibD10}.
Using the WKB method, it was shown that eq.~(\ref{eq1j}) is valid for half-odd-integer spin to leading order ($\mathcal{O} (n)$) \cite{bibcho}.
By employing the Teukolsky equation, which is applicable to both integer and half-odd-integer perturbations,
it was shown that eq.~(\ref{eq1j}) is valid for perturbations of general spin \cite{bibnov}.

Here, we derive the first-order correction to the asymptotic expression (\ref{eq1j}) thus extending the results for integer spin~\cite{bibx11}.
Remarkably, we arrive at the same expression as a function of the spin as in the bosonic case,
even though our starting point is different:
we employ the Teukolsky equation \cite{bibteu} whose solutions can be expressed in terms of confluent hypergeometric functions, whereas in the bosonic case discussed in \cite{bibx11}, the quasi-normal modes were obtained by solving the Regge-Wheeler equation \cite{bibRW} whose solutions are given in terms of Bessel functions.
In the case of massless Dirac perturbations, our results are in good agreement with numerical data \cite{bibD10,bibko}.


We are interested in the quasi-normal modes of massless perturbations of the Schwarzschild black hole whose metric is
\be ds^2 = - f(r) dt^2 + \frac{dr^2}{f(r)} + r^2 d\Omega_2^2 \ \ , \ \ \ \
f(r) = 1 - \frac{1}{r} \ee
where we have chosen units so that the black hole mass, radius of the horizon and Hawking temperature are respectively given by
\be\label{eqpara} M = \frac{1}{2}\ \ , \ \ \ \  r_0 = 1\ \ , \ \ \ \ T_H = \frac{1}{4\pi} \ee
Massless perturbations are described by the Teukolsky equation \cite{bibteu} for arbitrary spin, including both integer and half-odd-integer values.
This wave equation may be brought into a Schr\"odinger-like form~\cite{bibnov},
\be\label{eqwe}
\left( \frac{d^2}{dr_*^2}+\omega^2-V[r(r_*)] \right) \Psi(r_*) = 0
\ee
written in terms of the tortoise coordinate
\be\label{eqtort}
r_* = r+\ln(1-r)
\ee
The potential is given by
\be\label{eqpot}
V(r) = f(r)\left( \frac{\ell(\ell+1)}{r^2} + \frac{1}{r^3} \right)
+\frac{2i\omega j}{r} - \frac{3i\omega j}{r^2} + \frac{j^2}{4r^4}
\ee
for a spin-$j$ field of angular momentum $\ell$.

For quasi-normal modes, we demand the asymptotics
\be\label{eqasymp} \Psi(r_*) \sim e^{\pm i\omega r_*} \ \ , \ \ \ \ r_*\to \pm\infty \ee
so that the wave is outgoing at infinity ($r_*\to\infty$) and ingoing at the horizon ($r_*\to -\infty$).
More precisely, we shall define the boundary conditions in terms of the monodromy along a closed contour enclosing the black hole singularity and running counterclockwise, following \cite{bibx5}.
Such a contour may me deformed so that the monodromy receives contributions only from the singular points $r=1$ (the horizon) and infinity.
Near the horizon and at infinity, the wavefunction behaves respectively as
\be \Psi \sim (r-1)^{-i(\omega +ij/2)} \ \ , \ \ \ \ \Psi\sim r^{i(\omega -ij)} e^{i\omega r} \ee
The deformed contour runs counterclockwise at infinity and clockwise around $r=1$.
Therefore, the monodromy is
\be\label{eqmo} \mathcal{M} = e^{4\pi(\omega -ij/4)} \ee
Next, we deform the monodromy contour so it encircles the black hole singularity.
In this case, it receives a sole contribution from the rotation by $-2\pi$ near $r=0$.
We shall calculate the monodromy following \cite{bibnov}.
This will yield the asymptotic form of quasi-normal frequencies.
We shall then calculate the first-order correction to the asymptotic expression
and compare with numerical results \cite{bibD10,bibko}.

It is convenient to introduce the dimensionless coordinate $z = \omega r_*$.
In the complex $z$-plane, this rotation is of angle $-4\pi$, because for small $r$, $z\sim r^2$ on account of (\ref{eqtort}).
Expandng around the black hole singularity ($z = 0$),
the potential~(\ref{eqpot}) becomes
\be\label{eqVexp}
\frac{1}{\omega^2}V(z) = \frac{3ij}{2z} - \frac{4-j^2}{16z^2} + \frac{\mathcal{A}}{\omega^{1/2} z^{3/2}} + \mathcal{O} (1/\omega) \ \ , \ \ \ \
\mathcal{A} = \frac{\ell(\ell+1)+ \frac{1-j^2}{3}}{2\sqrt 2 } \ee
It is interesting to compare with the effective potential one obtains from the Regge-Wheeler equation \cite{bibRW} in the case of integer spin~\cite{bibx5}.
In the latter case, the leading term has a double pole at $z=0$, but the first-order correction is of the same form \cite{bibx11}.

The wave equation (\ref{eqwe}) may be written as
\be\label{eqwe1} \left( \mathcal{H}_0 + \frac{1}{\sqrt\omega} \mathcal{H}_1 + \dots \right) \Psi = 0 \ee
where
\be\label{eqHH} \mathcal{H}_0 \Psi = \frac{d^2\Psi}{dz^2} + \left[ 1 - \frac{3ij}{2z} - \frac{4-j^2}{16z^2} \right] \Psi \ \ , \ \ \ \ \mathcal{H}_1 = \frac{\mathcal{A}}{z^{3/2}} \ee
and we used the expansion (\ref{eqVexp}) of the potential.

We may solve (\ref{eqwe1}) perturbatively by expanding the wavefunction
\be\label{eqpsi1} \Psi = \Psi^{(0)} + \frac{1}{\sqrt\omega} \Psi^{(1)} + \dots \ee
The zeroth-order wave equation reads
\be \mathcal{H}_0 \Psi^{(0)} = 0 \ee
Its solutions are Whittaker functions. Two independent solutions are
\be\label{eqpsipm0} \Psi_\pm^{(0)} (z) = M_{\lambda,\pm\mu} (-2iz) \ \ , \ \ \ \
\lambda = \frac{3j}{4} \ \ , \ \ \mu = \frac{j}{4} \ee
These functions are not defined for even-integer spin. This can be remedied by adopting the Whittaker functions $W_{\lambda,\mu}(-2iz)$ and $W_{-\lambda,\mu} (-2ie^{-i\pi} z)$ as a basis instead.
The latter have more complicated rotation properties (around the origin), so we shall work with the basis (\ref{eqpsipm0}) for arbitrary spin and then
analytically continue the results to include the case of even-integer spin.

At large $z$, the wavefunctions (\ref{eqpsipm0}) behave respectively as
\be\label{eqasy} \Psi_\pm^{(0)} (z) \sim A_\pm (-2iz)^\lambda e^{iz} + B_\pm (-2ie^{-\pi i}z)^\lambda e^{-iz} \ee
where
\be\label{eqAB} A_\pm = \frac{\Gamma (1\pm 2\mu)}{\Gamma(\frac{1}{2} \pm\mu+\lambda)} e^{i\pi(\frac{1}{2} \pm\mu-\lambda)} \ \ , \ \ \ \ B_\pm = \frac{\Gamma (1\pm 2\mu)}{\Gamma(\frac{1}{2} \pm\mu-\lambda)} e^{-i\pi\lambda}\ee
In deriving the asymptotic form of the wavefunctions, we used the identity
\be\label{eqWid} M_{\lambda,\mu} (y) = \Gamma(1+2\mu) e^{-i\pi\lambda} \left[ \frac{1}{\Gamma(\frac{1}{2} +\mu-\lambda)} W_{-\lambda,\mu} (e^{-i\pi} y) + \frac{e^{i\pi(\frac{1}{2} +\mu)}}{\Gamma(\frac{1}{2} +\mu-\lambda)} W_{\lambda,\mu} (y) \right]\ee 
and the asymptotic form of the Whittaker function
\be W_{\lambda,\mu} (y) \sim y^\lambda e^{-\frac{y}{2}} \ee
Rotating around the $z=0$ singularity in the complex $z$-plane by $-4\pi$
changes the two wavefunctions (\ref{eqpsipm0}) by phases given by
\be\label{eq4pi} \Psi_\pm^{(0)} (e^{-4\pi i} z) = e^{\mp 4\pi i\mu} \Psi_\pm (z) \ee
where we used
\be M_{-\lambda,\mu} (e^{-i\pi} y) = e^{-i\pi (\frac{1}{2} + \mu)} M_{\lambda,\mu} (y) \ee
four times.

Let
\be M = \left( \begin{array}{cc} m_{11} & m_{12} \\ m_{21} & m_{22} \end{array} \right) \ee
be the monodromy matrix.
From the asymptotic form (\ref{eqasy}) and the rotation property (\ref{eq4pi}),
we deduce the eigenvalues and corresponding eigenvectors of the monodromy matrix,
\be M \left( \begin{array}{cc} A_+ \\ B_+ \end{array} \right) = e^{-4\pi i \mu} \left( \begin{array}{cc} A_+ \\ B_+ \end{array} \right) \ \ , \ \ \ \ M \left( \begin{array}{cc} A_- \\ B_- \end{array} \right) = e^{4\pi i \mu} \left( \begin{array}{cc} A_- \\ B_- \end{array} \right) \ee
They completely determine $M$. In particular,
\be\label{eqm11} m_{11} = \frac{ e^{-4\pi i \mu}A_+ B_- - e^{4\pi i \mu}A_- B_+}{A_+ B_- - A_- B_+} \ee
The other entries can also be found but are not needed for our purposes.
After some algebra, we obtain
\be\label{eqmo11} m_{11} = e^{i\pi (1-j)} (1+2\cos\pi j) \ee
$m_{11}$ should be equal to the monodromy (\ref{eqmo}) found using the contour that encircles the singularities at the horizon and infinity,
\be\label{eqmo12} m_{11} = \mathcal{M} \ee
This determines the spectrum of quasi-normal frequencies.
Using eqs.~(\ref{eqmo}) and (\ref{eqmo11}), we obtain the asymptotic form of the spectrum,
\be\label{eqsp0} \frac{\omega_n}{T_H} = -(2n+1) \pi i + \ln (1+2\cos\pi j) + \mathcal{O} (1/\sqrt n) \ee
where the Hawking temperature is given by (\ref{eqpara}).
This generalizes the result from the Regge-Wheeler equation for integer spin \cite{bibx5}.
For half-odd-integer spin (e.g., for a Dirac fermion), this reduces to
\be\label{eq0j} \frac{\omega_n}{T_H} = -(2n+1) \pi i  + \mathcal{O} (1/\sqrt n)\ \ , \ \ \ \ j+ \frac{1}{2} \in\mathbb{N} \ee
showing that asymptotically, the real part of the quasi-normal frequencies vanishes.

Next, we calculate the first-order correction to the asymptotic expression (\ref{eqsp0}) using the method of ref.~\cite{bibx11}.
It follows from eqs.~(\ref{eqwe1}) and (\ref{eqHH}) that the first-order correction to the wavefunction (\ref{eqpsi1}) obeys the wave equation
\be \mathcal{H}_0 \Psi^{(1)} = - \mathcal{H}_1 \Psi^{(0)} \ee
The corrections to the two basis vectors (\ref{eqpsipm0}) are given respectively by
\be\label{eqpsi2} \Psi_\pm^{(1)} (z) = a_\pm (z)\Psi_\pm^{(0)} (z) - b_\pm (z)\Psi_\mp^{(0)} (z) \ee
where we introduced the functions
\bea a_\pm (z) &=& \int_0^z \frac{dz'}{\mathcal{W}} \Psi_\pm^{(0)} (z') \mathcal{H}_1 \Psi_\mp^{(0)} (z') \nonumber\\
b_\pm (z) &=& \int_0^z \frac{dz'}{\mathcal{W}} \Psi_\pm^{(0)} (z') \mathcal{H}_1 \Psi_\pm^{(0)} (z') \eea
where $\mathcal{W} = 2\mu$ is the Wronskian.
Using the explicit expressions (\ref{eqpsipm0}) and (\ref{eqHH}), we may write
\bea\label{eq33} a_\pm (z) &=& \frac{\mathcal{A}}{4\mu} \int_0^z \frac{dz'}{{z'}^{3/2}} M_{\lambda,\pm\mu} (-2iz') M_{\lambda,\mp\mu} (-2iz') \nonumber\\
b_\pm (z) &=& \frac{\mathcal{A}}{4\mu} \int_0^z \frac{dz'}{{z'}^{3/2}} M_{\lambda,\pm\mu} (-2iz') M_{\lambda,\pm\mu} (-2iz') \eea
The asymptotic behavior (\ref{eqasy}) receives a first-order correction.
From eqs.~(\ref{eqpsi1}), (\ref{eqpsi2}) and the identity (\ref{eqWid}), we
deduce that for large $z$,
\be\label{eqasy1} \Psi_\pm (z) \sim A_\pm' (-2iz)^\lambda e^{iz} + B_\pm' (-2ie^{-\pi i}z)^\lambda e^{-iz} \ee
where
\be\label{eqABp} A_\pm' = \left( 1+\frac{1}{\sqrt\omega} \bar a_\pm \right) A_\pm - \frac{1}{\sqrt\omega}\bar b_\pm A_\mp \ \ , \ \ \ \ B_\pm' = \left( 1+\frac{1}{\sqrt\omega}\bar a_\pm \right) B_\pm - \frac{1}{\sqrt\omega}\bar b_\pm B_\mp \ee
and we defined
\be\label{eq36} \bar a_\pm = a_\pm (\infty) \ \ , \ \ \ \ \bar b_\pm = b_\pm (\infty) \ee
The monodromy (\ref{eqm11}) is corrected to
\be\label{eqm111} m_{11} \to \frac{ e^{-4\pi i \mu}A_+' B_-' - e^{4\pi i \mu}A_-' B_+'}{A_+' B_-' - A_-' B_+'} \ee
Using eq.~(\ref{eqABp}), this may be massaged to
\be\label{eqm1111} m_{11} \to m_{11} \left[ 1 -\frac{2i}{\sqrt\omega}\ \sin 4\pi\mu\ \frac{\bar b_+ A_- B_- + \bar b_- A_+ B_+}{e^{-4\pi i \mu}A_+ B_- - e^{4\pi i \mu}A_- B_+} + \mathcal{O} \left( \frac{1}{\omega} \right) \right]
\ee
Then the condition (\ref{eqmo12}) implies a modification to the asymptotic spectrum (\ref{eqsp0}) including first-order corrections,
\be\label{eqsp1} \frac{\omega_n}{T_H} = -(2n+1) \pi i + \ln (1+2\cos\pi j) -\frac{2i}{\sqrt{-in/2}}\ \sin 4\pi\mu\ \frac{\bar b_+ A_- B_- + \bar b_- A_+ B_+}{e^{-4\pi i \mu}A_+ B_- - e^{4\pi i \mu}A_- B_+} + \mathcal{O} (1/n) \ee
This is our main result. It should be compared with its counterpart from the Regge-Wheeler equation derived in \cite{bibx11} which is valid for integer spin $j$,
\bea\label{eqspint} \frac{\omega_n}{T_H} &=& -(2n+1) \pi i + \ln (1+2\cos\pi j)\nonumber\\
& & + \frac{1+i}{\sqrt n} \frac{\mathcal{A}}{4 \pi^{3/2}}\
\frac{\sin 2\pi j}{\sin \frac{3\pi j}{2}} \Gamma^2 \left( \frac{1}{4} \right) \Gamma \left( \frac{1}{4} + \frac{j}{2} \right) \Gamma \left( \frac{1}{4}- \frac{j}{2} \right) \nonumber\\
& & + \mathcal{O} (1/n) \ \ , \ \ \ \ j\in\mathbb{N} \eea
where $\mathcal{A}$ is given in (\ref{eqVexp}).

To obtain explicit expressions for the first-order spectrum (\ref{eqsp1}), let us concentrate on the case of half odd integer spin. In this case, there is a simplification due to the fact that
\be B_- = 0 \ \ , \ \ \ \ j+\frac{1}{2} \in\mathbb{N} \ee
as is evident from (\ref{eqAB}) for the parameters (\ref{eqpsipm0}).
Eq.~(\ref{eqpsi1}) simplifies to
\be\label{eqsp1a} \left. \frac{\omega_n}{T_H}\right|_{j+\frac{1}{2} \in\mathbb{N}} = -(2n+1) \pi i +\frac{2(1+i)}{\sqrt{n}}\ \frac{ A_+ }{ A_- }\ \bar b_- + \mathcal{O} (1/n) \ee
From eqs.~(\ref{eq33}) and (\ref{eq36}), we have
\be\label{eq36a} \bar b_- = \frac{\mathcal{A}}{4\mu} \int_0^\infty dz\ z^{-3/2} (M_{\lambda,-\mu} (-2iz))^2 \ee
For half odd integer spin,
the Whittaker function in (\ref{eq36a}) may be written in terms of a Laguerre polynomial as
\be M_{\lambda,-\mu} (y) = \frac{\Gamma(j+1/2)\Gamma(1-j/2)}{\Gamma(j/2+1/2)} y^{1/2-j/4} e^{-y/2} L_{j-1/2}^{-j/2} (y) \ \ , \ \ \ \ j+ \frac{1}{2} \in\mathbb{N} \ee
which simplifies the calculation of the integral (\ref{eq36a}).
However, one ought to be cautious with such simplifications, because in general these integrals may only be defined by analytic continuation to the desired spin $j$.

\begin{figure}
\begin{center}
\includegraphics[width=6.5in]{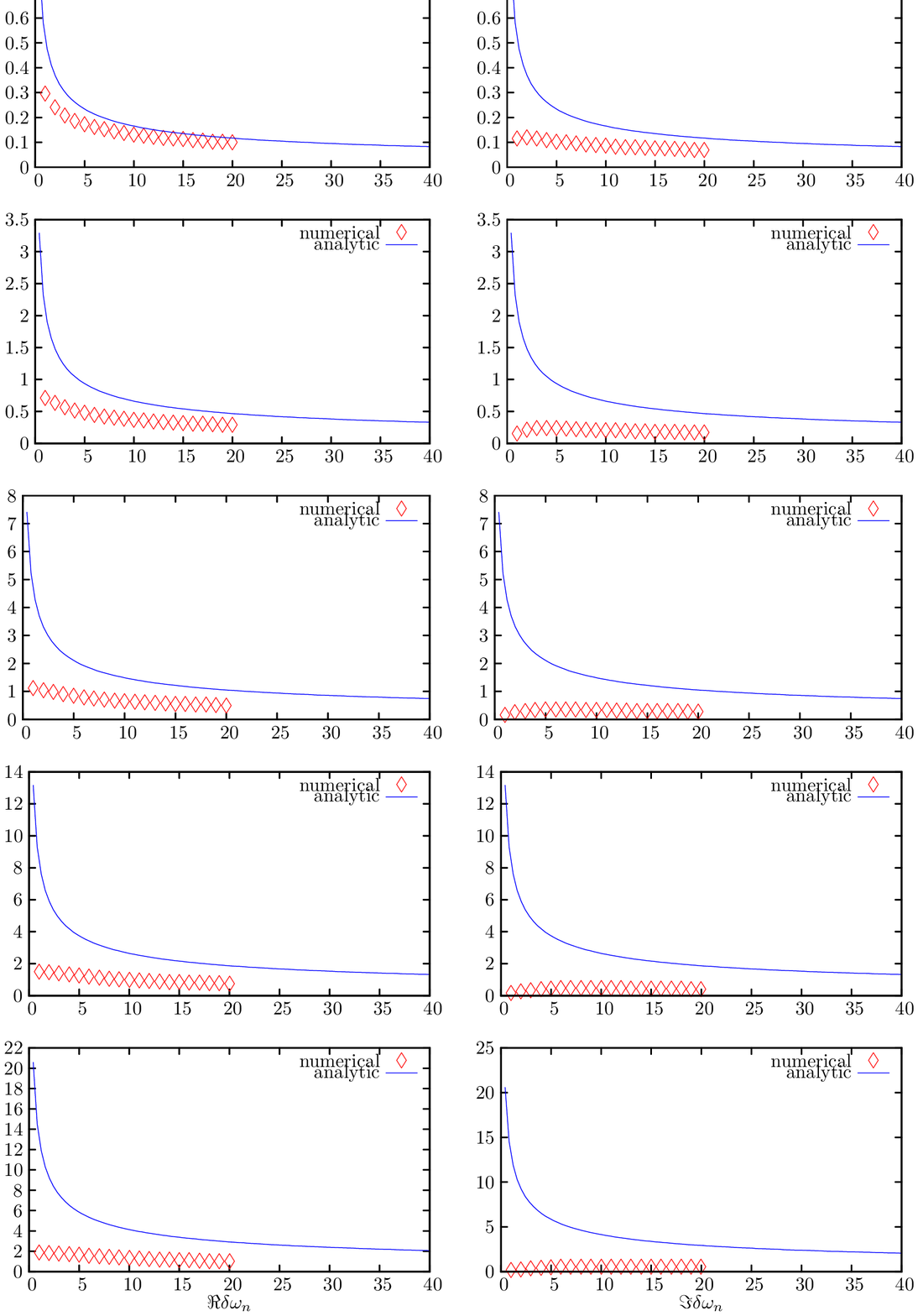}
\end{center}
\vspace{-1.25in}
\caption{\label{fig1}Quasi-normal frequencies of massless Dirac fermions for various values of angular momentum (from top to bottom: $\ell+j = 1,2,3,4,5$); solid lines are graphs of our analytic expression~(\ref{eqdelom}); diamonds represent numerical data~\cite{bibD10}.}
\end{figure}

For $j=1/2$ (Dirac fermion), we obtain from (\ref{eq36a})
\bea \bar b_- &=& \frac{\mathcal{A}}{j}
(-2i)^{3/2}\int_0^\infty dz\ (-2iz)^{-\frac{1}{2}-\frac{j}{2}} e^{2iz} +\dots
\nonumber\\
&=& \frac{\mathcal{A}}{j}
(-2i)^{1/2} \Gamma \left( \frac{1}{2} - \frac{j}{2} \right) + \dots\eea
where the dots represent terms that vanish as $j\to \frac{1}{2}$.
In the limit $j= \frac{1}{2}$, we obtain
\be\label{eq36b} \bar b_-\Big|_{j= \frac{1}{2}} = 2\mathcal{A} (-2i)^{1/2} \Gamma\left( \frac{1}{4} \right) \ee
Using eqs.~(\ref{eqVexp}), (\ref{eqAB}), (\ref{eqsp1a}) and (\ref{eq36b}), we deduce the spectrum of a massless Dirac fermion,
\be\label{eqsp1aD} \left. \frac{\omega_n}{T_H}\right|_{j=\frac{1}{2}} = -(2n+1) \pi i +\frac{1+i}{2\sqrt{n}}\ \left( \ell + \frac{1}{2} \right)^2\ \Gamma^2\left( \frac{1}{4} \right)+ \mathcal{O} (1/n) \ee
To compare with numerical results \cite{bibD10,bibko}, we shall work with the first-order correction
\be\label{eqdelom} \left. \delta\omega_n\right|_{j=\frac{1}{2}} = \frac{1+i}{8\pi\sqrt{n}}\ \left( \ell + \frac{1}{2} \right)^2\ \Gamma^2\left( \frac{1}{4} \right) \ee
which is the difference between the first-order (\ref{eqsp1aD}) and zeroth-order
(\ref{eq0j}) expressions multiplied by the Hawking temperature (\ref{eqpara}).
In fig.~\ref{fig1}, we compare our analytic expression with the numerical results of ref.~\cite{bibD10}.
We obtain fairly good agreement with the discrepancy between numerical and analytic results increasing with angular momentum. This is because the first-order correction to the asymptotic form (\ref{eq0j}) is proportional to the square of the angular momentum quantum number $\ell+j$.
Moreover, it should be noted that the corrections are $\mathcal{O} (1/\sqrt n)$, so they become significant for overtones with $n\lesssim 100$.

\begin{figure}
\begin{center}
\includegraphics{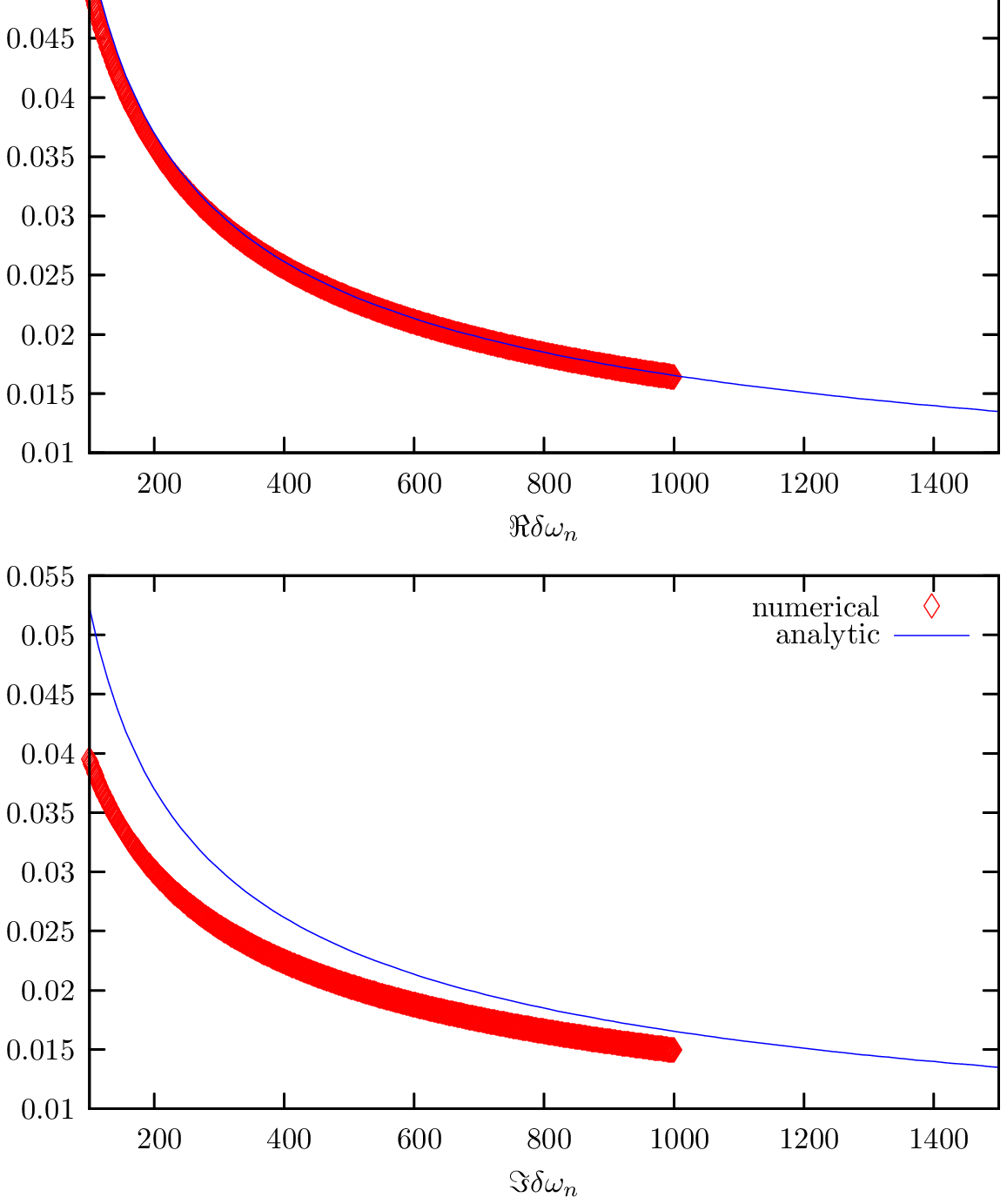}
\end{center}
\vspace{-4.25in}
\caption{\label{fig2}High overtones ($n\ge 100$) of massless Dirac fermions for $\ell+j = 1$: solid lines are graphs of our analytic expression~(\ref{eqdelom}); diamonds represent numerical data~\cite{bibko}.}
\end{figure}

\begin{figure}
\begin{center}
\includegraphics{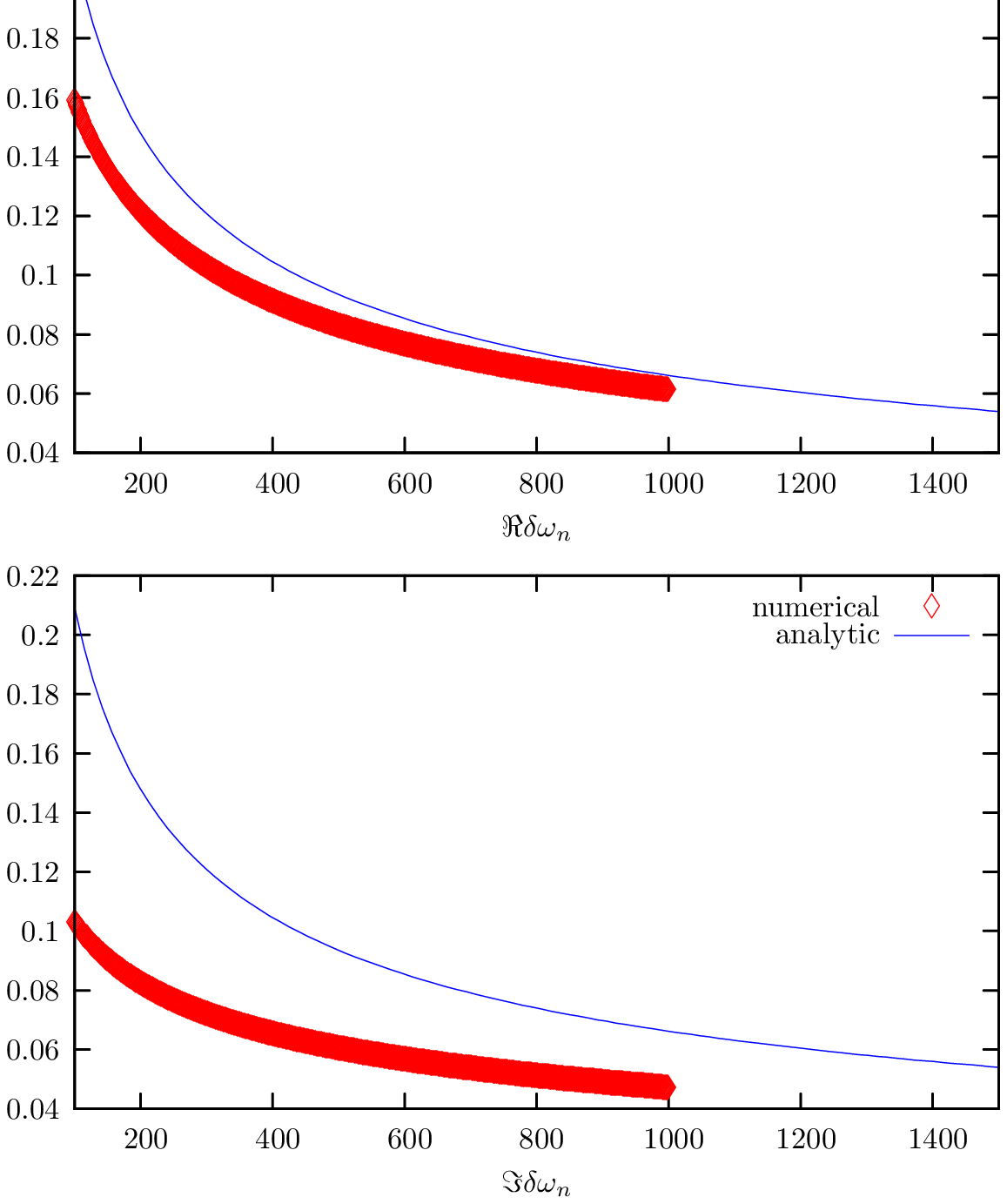}
\end{center}
\vspace{-4.25in}
\caption{\label{fig3}High overtones ($n\ge 100$) of massless Dirac fermions for $\ell+j = 2$: solid lines are graphs of our analytic expression~(\ref{eqdelom}); diamonds represent numerical data~\cite{bibko}.}
\end{figure}

In figures \ref{fig2} and \ref{fig3} we compare our expression (\ref{eqdelom}) with numerical results for high overtones ($n\ge 100$) \cite{bibko} for $\ell+ j = 1,2$, respectively.
The agreement improves greatly, as expected. In the case of lowest angular momentum ($\ell+j = 1$), the agreement is excellent (fig.~\ref{fig2}).

For a massless field of spin $j=3/2$, we obtain from (\ref{eq36a})
\bea \bar b_- &=& \frac{\mathcal{A}}{j}(-2i)^{3/2}
\int_0^\infty dz\ (-2iz)^{-\frac{1}{2}-\frac{j}{2}} e^{2iz} \left( 1 + \frac{2iz}{1-j/2} + \dots \right)^2 \nonumber\\
&=& \frac{\mathcal{A}}{j}(-2i)^{1/2} \frac{(\frac{3}{2} -j)\Gamma\left( \frac{1}{2} -\frac{j}{2} \right)}{2(1-j/2)^2} +\dots \eea
where we once again omitted terms which vanish as $j\to \frac{3}{2}$.
Evidently, in the limit $j\to \frac{3}{2}$,
\be\label{eq36c} \bar b_-\Big|_{j= \frac{3}{2}} = 0 \ee
showing that there are no $\mathcal{O} (1/\sqrt n)$ corrections to the spectrum,
\be\label{eqsp1b} \left. \frac{\omega_n}{T_H}\right|_{j=\frac{3}{2}} = -(2n+1) \pi i + \mathcal{O} (1/n) \ee
For a massless field of spin $j=5/2$, we obtain from (\ref{eq36a})
\bea \bar b_- &=& \frac{\mathcal{A}}{j}(-2i)^{3/2}
\int_0^\infty dz\ (-2iz)^{-\frac{1}{2}-\frac{j}{2}} e^{2iz} \left( 1 + \frac{4iz}{1-j/2} - \frac{4z^2}{(1-j/2)(2-j/2)} +\dots\right)^2 \nonumber\\
&=& \frac{\mathcal{A}}{j}(-2i)^{1/2} \frac{(41-40j+8j^2)\Gamma\left( \frac{1}{2} -\frac{j}{2} \right)}{16(1-j/2)(2-j/2)} +\dots \eea
In the limit $j\to 5/2$, we obtain
\be\label{eq36d} \bar b_-\Big|_{j= \frac{5}{2}} = \frac{128}{15}\mathcal{A} (-2i)^{1/2} \ \Gamma\left( \frac{1}{4} \right) \ee
Using eqs.~(\ref{eqAB}), (\ref{eqsp1a}) and (\ref{eq36d}), we deduce the spectrum of a massless fermion of spin $j=5/2$,
\be\label{eqsp1c} \left. \frac{\omega_n}{T_H}\right|_{j=\frac{5}{2}} = -(2n+1) \pi i +\frac{1+i}{\sqrt{2n}}\ \mathcal{A} \ \Gamma^2\left( \frac{1}{4} \right)+ \mathcal{O} (1/n) \ee
where $\mathcal{A}$ is given in (\ref{eqVexp}).

It is interesting to note that all spectra derived above for fields of half odd integer spin
may also be obtained by analytically continuing the general expression (\ref{eqspint}) which was obtained for integer spins \cite{bibx11}.
All cases we checked agree with (\ref{eqspint}) but we have been unable to find a general argument applicable to arbitrary half odd integer spin.
The same is true for integer spin spectra derived from the Teukolsky equation, although the calculation is more cumbersome (except for zero spin, in which case the Teukolsky equation reduces to the Regge-Wheeler equation).

In conclusion, we have derived analytic expressions for quasi-normal frequencies of perturbations of Schwarzschild black holes of arbitrary spin including both bosonic and fermionic fields. Our results were in good agreement with numerical data \cite{bibD10,bibko}.
Our calculation was based on the monodromy argument of ref.~\cite{bibx5} extended to the Teukolsky equation \cite{bibnov}.
We derived the asymptotic spectrum including first-order corrections extending the argument in \cite{bibx11}.
We obtained explicit expressions for various spins and found that they were in agreement with the analytic expression derived in \cite{bibx11} for bosonic fields (after a simple analytic continuation of the spin to non-integer values).
This is remarkable given that our starting point was the Teukolsky equation whereas the bosonic spectrum was derived in \cite{bibx11} starting from the Regge-Wheeler equation.
It would be interesting to further investigate the relation between the two equations and understand the range of applicability of the Teukolsky equation
as the latter offers significant calculational advantages.

\section*{Acknowledgments}
We wish to thank Fu-Wen Shu for useful discussions and R.~A.~Konoplya for sending us his results.
G.~S.~was supported in part by the US Department of Energy under grant DE-FG05-91ER40627.
The work of S.~M.~was partially supported by the Thailand Research
Fund.
S.~M.~also gratefully acknowledges the hospitality of the Department of Physics and
Astronomy at the University of Tennessee where part of the work was performed.

\newpage

\begin{thebibliography}{99}

\bibitem{bibx1} O.~Dreyer, \prl{90} (2003) 081301; {\tt gc-qr/0211076}.
\bibitem{bibx2} G.~Kunstatter, \prl{90} (2003) 161301; {\tt gr-qc/0212014}.
\bibitem{bibx3} L.~Motl, {\sl Adv.~Theor.~Math.~Phys.~\bf 6} (2003) 1135;
{\tt gr-qc/0212096}.
\bibitem{bibx4} A.~Corichi, \pr{67} (2003) 087502; {\tt gr-qc/0212126}.
\bibitem{bibx5} L.~Motl and A.~Neitzke, {\sl Adv.~Theor.~Math.~Phys.~\bf 7} (2003) 2; {\tt hep-th/0301173}.
\bibitem{bibx6} E.~Berti and K.~D.~Kokkotas, \pr{68} (2003) 044027; {\tt hep-th/0303029}.
\bibitem{bibx7} R.~A.~Konoplya, \pr{68} (2003) 024018; {\tt gr-qc/0303052}.
\bibitem{bibx8} A.~Maassen van den Brink, {\sl J.~Math.~Phys.~\bf 45} (2004) 327; {\tt gr-qc/0303095}.
\bibitem{bibx9} A.~P.~Polychronakos, \pr{69} (2004) 044010; {\tt hep-th/0304135}.
\bibitem{bibx10} E.~Berti, V.~Cardoso, K.~Kokkotas and H.~Onozawa, \pr{68} (2003) 124018; {\tt hep-th/0307013}.
\bibitem{bibx11} S.~Musiri and G.~Siopsis, \cqg{20} (2003) L285; {\tt hep-th/0308168}.
\bibitem{bibbi1} J.~F.~Barbero G., \pr{51} (1995) 5507; {\tt gr-qc/9410014}.
\bibitem{bibbi2} G.~Immirzi, {\sl Nucl.~Phys.~Proc.~Suppl.~\bf 57} (1997) 65;
{\tt gr-qc/9701052}.
\bibitem{biblqg1} C.~Rovelli and L.~Smolin, \np{442} (1995) 593;
{\tt gr-qc/9411005}.
\bibitem{biblqg2} A.~Ashtekar and J.~Lewandowski, \cqg{14} (1997) A55;
{\tt gr-qc/9602046}.
\bibitem{biblqg3} C.~Rovelli, {\sl Living Rev.~Rel.~\bf 1} (1998) 1;
{\tt gr-qc/9710008}.
\bibitem{biblqg4} T.~Thiemann, {\tt gr-qc/0110034}.
\bibitem{biblqg5} C.~Rovelli and P.~Upadhya, {\tt gr-qc/9806079}.
\bibitem{bibn1} S.~Chandrasekhar and S.~Detweiler, {\sl Proc.~R.~Soc.~London,
Ser.~A \bf 344} (1975) 441.
\bibitem{bibn2} E.~W.~Leaver, {\sl Proc.~R.~Soc.~London,
Ser.~A \bf 402} (1985) 285.
\bibitem{bibn3} H.~P.~Nollert, \pr{47} (1993) 5253.
\bibitem{bibn4} N.~Andersson, \cqg{10} (1993) L61.
\bibitem{bibn5} A.~Bachelot and A.~Motet-Bachelot, {\sl Annales Poincar\`e
Phys.~Theor.~\bf 59} (1993) 3.
\bibitem{bibhod} S.~Hod, \prl{81} (1998) 4293; {\tt gr-qc/9812002}.
\bibitem{bibbm} J.~D.~Bekenstein and V.~F.~Mukhanov, \pl{360} (1995) 7; {\tt gr-qc/9505012}.
\bibitem{bibD1} S.~Iyer and C.~M.~Will, \pr{35} (1987) 3621.
\bibitem{bibD2} V.~Cardoso and J.~P.~S.~Lemos, \pr{63} (2001) 124015; {\tt gr-qc/0101052}.
\bibitem{bibD3} H.~T.~Cho, \pr{68} (2003) 024003; {\tt gr-qc/0303078}.
\bibitem{bibD4} R.~A.~Konoplya, \pr{68} (2003) 024018; {\tt gr-qc/0303052}.
\bibitem{bibD5} A.~Zhidenko, \cqg{21} (2004) 273; {\tt gr-qc/0307012}.
\bibitem{bibD6} W.~Zhou and J.-Y.~Zhu, {\sl Int.~J.~Mod.~Phys.~\bf D13} (2004) 1105; {\tt gr-qc/0309071}.
\bibitem{bibD7} J.~Jing, \pr{69} (2004) 084009; {\tt gr-qc/0312079}.
\bibitem{bibD8} K.~H.~C.~Castello-Branco, R.~A.~Konoplya and A.~Zhidenko, \pr{71} (2005) 047502; {\tt hep-th/0411055}.
\bibitem{bibD9} H.~T.~Cho, \cqg{22} (2005) 775; {\tt gr-qc/0411090}.
\bibitem{bibD10} J.~Jing, \pr{71} (2005) 124006; {\tt gr-qc/0502023}.
\bibitem{bibcho} H.~T.~Cho, \pr{73} (2006) 024019; {\tt gr-qc/0512052}.
\bibitem{bibnov} I.~B.~Khriplovich and G.~Yu.~Ruban, {\sl Int.~J.~Mod.~Phys.~\bf D15} (2006) 879; {\tt gr-qc/0511056}.
\bibitem{bibteu} S.~A.~Teukolsky, \prl{29} (1972) 1114;\newline
S.~A.~Teukolsky, {\sl Astrophys.~J.~\bf 185} (1973) 635;\newline
R.~G\"uven, \pr{22} (1980) 2327.
\bibitem{bibRW} T.~Regge and J.~A.~Wheeler, \pr{108} (1957) 1063.\newline
F.~J.~Zerilli, \pr{2} (1970) 2141.
\bibitem{bibko} R.~A.~Konoplya, private communication.
\end{thebibliography}
\end{document}